\documentclass[prb,twocolumn,showpacs,amsmath,amssymb,preprintnumbers,superscriptaddress]{revtex4}
\usepackage{dcolumn}
\usepackage{bm}
\usepackage{graphicx}
\usepackage{subfigure}
\usepackage{color}
\usepackage{scalerel,amssymb}

\begin{document}

\title{Critical behavior and magnetocaloric effect in ferromagnetic nano-crystalline Pr$_2$CoMnO$_6$}

\author{Ilyas Noor Bhatti}\email{inoorbhatti@gmail.com}\affiliation{Department of Physics, Jamia Millia Islamia University, New Delhi - 110025, India.}
\author{Imtiaz Noor Bhatti}\affiliation{School of Physical Sciences, Jawaharlal Nehru University, New Delhi - 110067, India.}
\author{Rabindra Nath Mahato}\affiliation{School of Physical Sciences, Jawaharlal Nehru University, New Delhi - 110067, India.}
\author{M. A. H. Ahsan}\affiliation{Department of Physics, Jamia Millia Islamia University, New Delhi - 110025, India.}

\begin{abstract}
To understand the nature of magnetic phase transition in nano-crystalline Pr$_2$CoMnO$_6$, in present study we have investigated the critical behavior and magnetocaloric effect. To estimate the critical exponents, various methods have been adopted like; the modified Arrott plots (MAP), the Kouvel-Fisher method (KF) and the critical isotherm analysis. This material shows a second order type a paramagnetic (PM) to ferromagnetic (FM) phase transition around 160 K. The critical exponents obtained from  modified Arrott plots are $\beta$ = 0.531, $\gamma$ = 0.935 and $T_c$ = 160 K. Kouvel-Fisher method gives the exponents as $\beta$ = 0.533$\pm$0.001 with $T_c$ = 160.72$\pm$0.03 K and  $\gamma$ = 0.932$\pm$0.003 with $T_c$ = 160.15$\pm$0.05 K. The third exponent $\delta$ = 2.763$\pm$0.005 obtained from critical isothermal is in agreement with Wisdom scaling rule. The estimated critical exponents do not exactly match with any established universality class, however, the deduced exponent values suggest that spin interaction in the present material is close to mean-field model which suggests the existence of long-range ferromagnetic order in nano-crystalline Pr$_2$CoMnO$_6$. The reliability of critical exponents is checked by universal scaling hypothesis on magnetic data across T$_c$. We have calculated the magnetic entropy change from magnetic data and found maximum value of -$\Delta$S = 2.05 (J kg$^{-1}$ K$^{-1}$) for 50 kOe at 180 K. Moreover, field dependent change in magnetic entropy obeys scaling and also indicates that the magnetic interaction is close to mean-field type.
\end{abstract}

\maketitle
\section{Introduction}
 Double-perovskite oxides with general formula A$_2$BB$^{\prime}$O$_6$  (A is an alkaline-earth or rare-earth metal ion; B and B$^{\prime}$ are transition metals ions) receives significant interest in condanced matter physics because of the wide range of physical properties, ranging from large magnetocaloric effects (MCEs),\cite{moon, mahato} multiferroicity,\cite{syv, kumar} magnetodielectric behaviour,\cite{singh, gua, uma, yi} pyroelectric effect,\cite{sco1, sco2} ferroelectricity\cite{zli, jsu} etc. Among the family of double perovskite oxides 3$d$ base double perovskit received much attention, since these material shows potential for exotic physical properties.\cite{rov, dass, son, dya, choudhury, kawa, lil, bos} These physical properties are largely decided by the exchange interaction between B and B$^{\prime}$ cations. These cations have different oxidation state and can be found in low- or high- spin state depending upon crystal field energy,\cite{ala} thus coupled differently. The spin state and interaction among the cations via oxygen i.e. B-O-B$^{\prime}$ decides the physical properties specially the magnetic properties in these compounds.\cite{maya} For instance, La$_2$CoMnO$_6$ for perfectly order system Co$^{2+}$-O-Mn$^{4+}$ will involve in ferromagnetic ordering results from strong exchange interaction. In case if there is an anti-site disorder created in the system the Co$^{2+}$-O-Co$^{2+}$ and Mn$^{4+}$-O-Mn$^{4+}$ will involve in super-exchange interaction and will give rise to antiferromagnetic ordering.\cite{lshi, shi, liu} Furthermore, if Co$^{3+}$ and Mn$^{3+}$ valency is created in the system in the form of disorder then a new ferromagnetic transition occur at 150 K.\cite{shi} Compounds with Co/Mn at B/B$^{\prime}$ have shown some interesting physical phenomenon for instance, Gd$_2$CoMnO$_6$ is a ferromagnetic material which at low temperature suffer antiferromagnetic transition and at very low temperature spin glass state have also been reported in this material. Further, Gd$_2$CoMnO$_6$ show very large magnetocaloric effect at low temperature where maximum entropy change is around 35 J kg$^{-1}$ K$^{-1}$.\cite{wang} Recently, pyroelectric effect has been observed in ferromagnetic (Er,Y)$_2$CoMnO$_6$ which is believed to be due to thermal induced depolarization current.\cite{sco1, sco2} Nano-crystalline Pr$_2$CoMnO$_6$ is  ferromagnetic insulator which shows large magnetoresistance and magnetodielectric effect below magnetic transition. This compound also feature Griffith phase singularity well above T$_c$ in paramagnetic state.\cite{ilyas, khemchand}

However, in most of these compounds the nature of magnetic phase transition and exchange interaction need to be addressed to understand the underlying physics in further detail. The investigation on nature of magnetic transition in magnetic materials gives vital information about the thermodynamics of transition.\cite{zhang} In materials which exhibit second order phase transition i.e. where the phase transition takes place continuously over the critical region. It is a well established fact that in critical region the physical parameters obey power law. Where in magnetic materials magnetic moment and susceptibility are order parameters in the asymptotic region of phase transition. The order parameter and range of interaction obey power law behavior across phase transition and can be characterized by various critical exponents and scaling behavior. The critical analysis and scaling for material which exhibit second order type paramagnetic to ferromagnetic phase transition can provide vital information about the nature of magnetic state and exchange coupling. \cite{michael, stanley}

In this paper, we present the investigation of critical analysis and magnetic entropy change across PM to FM phase transition in nano-crystalline Pr$_2$CoMnO$_6$. We have evaluated the critical exponents in asymptotic region near T$_c$ using various rigorous methods. The determined critical exponents for the nano-crystalline Pr$_2$CoMnO$_6$ closely matches with the mean-field  interaction model with long range magnetic exchange coupling. We further studied the change in magnetic entropy $\Delta$S, where the field dependency of $\Delta$$S_M$ follows a power law behavior with scaled field i.e  $\Delta$$S_M$$\propto$H$^(2/3)$, which agrees with the prediction of meam field theory in the vicinity of phase transition close to $T_c$. We found that the maximum value of entropy change is -$\Delta$$S_M^max$(T) = 2.05 J kg$^{-1}$ K$^{-1}$ for an applied field of 50 kOe. We conclude that the critical exponents determined from different self-consistent methods, the scaling hypothesis and magnetic entropy change exhibits that the nano-crystalline Pr$_2$CoMnO$_6$ close to the mean field interaction model.

\section{Methods}
\subsection{Experimental details}
The single-phase nano-crystalline sample has been prepared using sol-get method. Details of sample preparation and characterization are given elsewhere.\cite{ilyas, khemchand}  The magnetic properties of nano-crystalline Pr$_2$CoMnO$_6$ is studied by magnetization measurements, where magnetization ($M$) data as a function of temperature ($T$) and magnetic field ($H$) have been collected using physical properties measurement system (PPMS) by Cryogenic Inc. To locate the PM to FM phase transition, temperature dependent magnetization M(T) is measure in the temperature range of 2 K to 300 K in an applied field of 0.5 kOe. In order to carry out critical analysis in the vicinity of PM to FM phase transition magnetization data is required in the asymptotic region of second order type PM to FM phase transition. Thus, magnetic isotherms ($M(H)$) are collected across $T_c$ in temperature range from 148 to 176 K with a temperature step $\Delta T$ = 2 K. To calculate the change in magnetic entropy $\Delta S_M$, the $M(H)$ data have been collected in wide temperature range 110 to 250 K up to field 50 kOe.

\subsection{Scaling Analysis}
 According to scaling hypothesis, across second order phase transition the critical exponent $\beta$ associated with the spontaneous magnetization ($M_S$) below $T_c$, $\gamma$ associated with initial inverse magnetic susceptibility ($\chi^{-1}$) above $T_c$ and $\delta$ associated with the magnetization at $T_c$ follow power-law behavior with temperature as given below:\cite{{stanley}}

\begin{eqnarray}
M_S(T) = M_0(-\epsilon)^\beta, &\epsilon < 0
\end{eqnarray}

\begin{eqnarray}
\chi_0^{-1}(T) = \Gamma(\epsilon)^\gamma, &\epsilon > 0
\end{eqnarray}

\begin{eqnarray}
M = X(H)^{1/\delta}, &\epsilon = 0
\end{eqnarray}

where the reduced temperature is represented by $\epsilon$ = ($T$ - $T_c$)/$T_c$. The $M_0$, $\Gamma$ and $X$ are the critical amplitudes and $\beta$, $\gamma$ and $\delta$ are the critical exponents. 

In the asymptotic critical region, the critical exponents should follow scaling hypothesis. According to this magnetic equation of state, the magnetization $(M(H,\epsilon))$, the magnetic field $(H)$ and the temperature $(T)$ obey following relationship,\cite{stanley}

\begin{eqnarray}
	M(H, \epsilon) = \epsilon^{\beta} f_\pm \left( \frac{H}{\epsilon^{\beta + \gamma}}\right)
\end{eqnarray}

where the $f_+$ for $(T > T_c)$ and $f_-$ for $(T < T_c)$ are defined as the regular functions. Eq. 4 implies that  renormalized magnetization $m$ = $M(H,\epsilon)$$\epsilon^{-\beta}$  plotted as function of renormalized field $h$ = $H$$\epsilon^{-(\beta + \gamma)}$ with correct set of exponents and $T_c$ will generate two universal curves one for temperature above $T_c$ and another for temperature below $T_c$.

\begin{figure}[t]
	\centering
		\includegraphics[width=8cm]{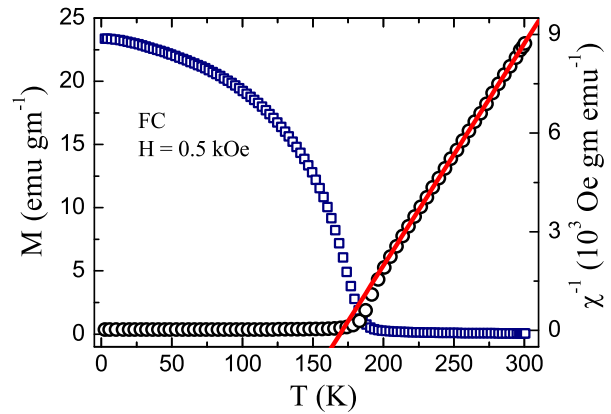}
\caption{(Color online) Temperature dependent magnetization ($M(T)$) measured under FC protocol in applied field of 0.5 kOe (left axis)  where as same data is plotted as inverse susceptibility (right axis) for nano-crystalline Pr$_2$CoMnO$_6$. Solid line is fitting due to Curie Weiss law.}
	\label{fig:Fig1}
\end{figure}

\begin{figure*}[t]
	\centering
		\includegraphics[width=18cm]{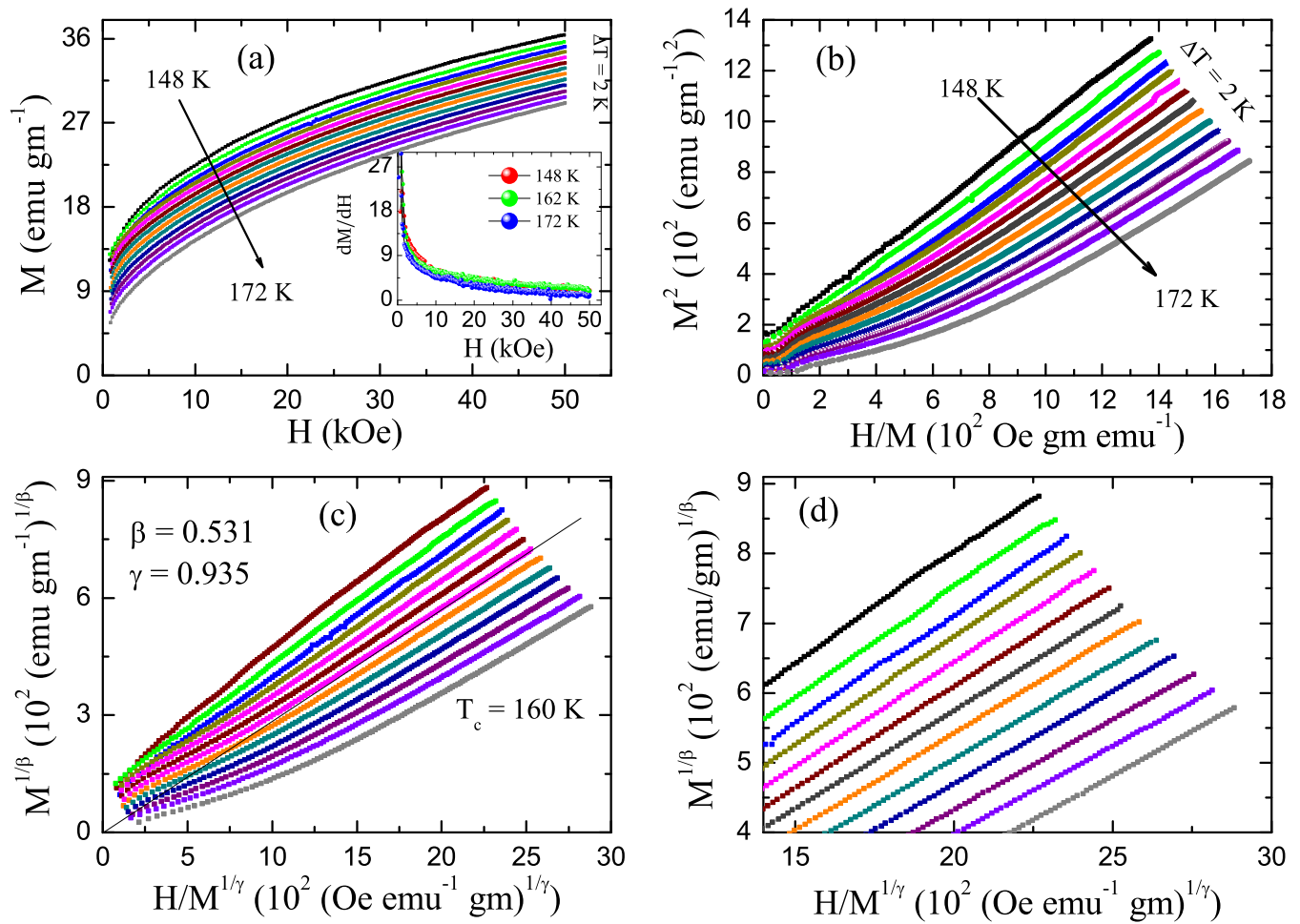}
	\caption{(Color online) (a) Magnetic isotherms ($M(H)$) measured at different temperatures across $T_c$ for nano-crystalline Pr$_2$CoMnO$_6$ are shown. Inset shows derivations of $M(H)$ data at selective temperatures. (b) Arrott plot ($M^2$ vs $H/M$) as obtained from Fig 2a. (c) Modified Arrott plot (Eq. 5) constructed from isotherms is shown with exponents $\beta$ = 0.531 and $\gamma$ = 0.935 for nano-crystalline Pr$_2$CoMnO$_6$. (d) Shows the modified Arrott plot zoomed at higher field for clarity.}
	\label{fig:Fig2}
\end{figure*}

\section{Results and Discussions}
\subsection{Critical Analysis}
Fig. 1 (left axis) shows the temperature dependent magnetization data $M(T)$ collected following field cooled. (FC) protocol in applied field of 0.1 kOe. It is clear from the figure that with decreasing temperature, magnetization suddenly began to rise below 190 K which is marked by PM to FM phase transition. The Magnetization data is plotted in terms of inverse susceptibility ($\chi^{-1}$) against temperature shown in Fig. 1 (right axis). The magnetic susceptibility follows Curie Weiss behavior above observed paramagnetic temperature $\sim$172.67 K above 206 K marked as T$_G$ below which Griffiths phase appear as reported in our previous work till T$_c$.\cite{ilyas} Nano-crystalline Pr$_2$CoMnO$_6$ undergoes a PM to FM phase transition which is evident and confirmed from positive slope of M(H) data around T$_c$ obey Benerjee criteria.\cite{bee} This implies that critical analysis can be carried out in critical region of phase transition in this material. We have investigated critical exponents across T$_c$ for better understanding of magnetic behavior and phase transition.

To understand the nature of magnetic state in nano-crystalline Pr$_2$CoMnO$_6$ we have  evaluated the exponents by measuring magnetic isotherms $M(H)$ around the PM to FM phase transition temperature $T_c$. The exponents have been determined by adopting different methods which includes, Arrott plot,\cite{arrott}, Kouvel-Fisher analysis \cite{kf} and critical isotherm analysis. Fig. 2a shows $M$ vs $H$ plots for nano-crystalline Pr$_2$CoMnO$_6$ measured in temperature range from 148 to 172 K at an interval of $\delta$T = 2 K. Inset Fig. 2a shows the decreasing slope of $dM/dH$ vs $H$ which implies a second-order type PM to FM phase transition in this material. Fig. 2b shows the isothermal magnetization $M(H)$ data plotted in terms of Arrott plot\cite{arrott} i.e., $M^2$ vs $H/M$ where positive slope of $H/M$ vs $M^2$ plot further satisfies the Benerjee criteria for second order phase transition. Generally, the magnetic systems for which the Arrott plot forms a set of parallel straight lines at high field regime is believed to follows mean-field spin interaction model. $M_S$ and $\chi_0$ are obtained from the intercept on positive $M^2$ and $H/M$ axis respectively, while extrapolating the high field regime in Arrott plot. Moreover, the isotherm which passes through origin in Arrott plot, known as critical isothermal gives $T_c$ of the material. Fig. 2b clearly shows that  isotherm curves are not perfectly parallel straight lines, however shows little deviation from which suggest that to make these curves into parallel straight lines minor tuning of exponents $\beta$ and $\gamma$ is required.  Further, it is worth noting here that the exponents are close to mean-field interaction model. 

Fig. 2c shows the isothermal magnetization data measured at different temperatures $M(T, H)$ close to $T_c$ is plotted using generalized modified Arrott plotts given by the following equation of state:\cite{arrott1}

\begin{eqnarray}
	\left(\frac{H}{M}\right)^{1/\gamma} = a \frac{T - T_c}{T_c} + b M^{1/\beta}
\end{eqnarray}

where the $a$ and $b$ are the constants. Fig. 2c is used to determine the value of spontaneous magnetization $M_S$(T, 0) and initial inverse susceptibility $\chi_0^{-1}$(T). Eq. 5 contains two unknown parameters $\beta$ and $\gamma$. Through careful exercise of tuning parameter $\beta$ and $\gamma$ parallel straight lines can be obtained. Here, we have followed the rigorous iterative process as described in Bhatti \textit{etal}.\cite{bhatti} The Fig. 2c shows modified Arrott plot with the so obtained exponents $\beta$ = 0.531 and $\gamma$ = 0.935. It is evident in the figure that a set of parallel straight lines are obtained in higher field regime. Fig. 2d shows the modified Arrott plot zoomed at high field regime for clarity. The lines in lower field regime, however, are curved due to different demagnetization factor involved at low field. We find that isotherm taken at temperature 160 K passes through origin which is the critical isothermal and gives the $T_c$ of this material.

\begin{figure}
	\centering
		\includegraphics[width=8cm]{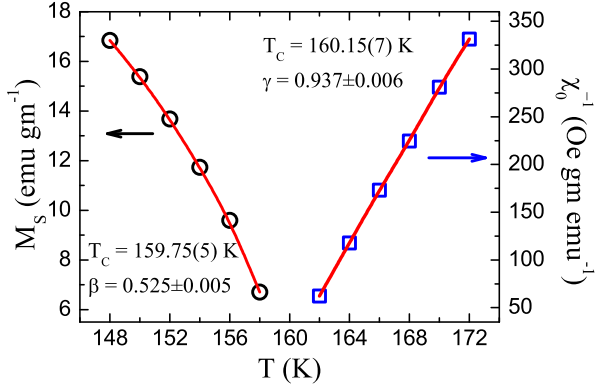}
	\caption{(Color online) Spontaneous magnetization $M_S$ (left axis) and inverse initial susceptibility $\chi_0^{-1}$ (right axis) as determined from the high field extrapolation of modified Arrott plot in Fig. 2c plotted as a function of temperature are shown for nano-crystalline Pr$_2$CoMnO$_6$. Solid lines are due to fitting with Eqs. 1 and 2.}
	\label{fig:Fig3}
\end{figure}

\begin{figure}
	\centering
		\includegraphics[width=8cm]{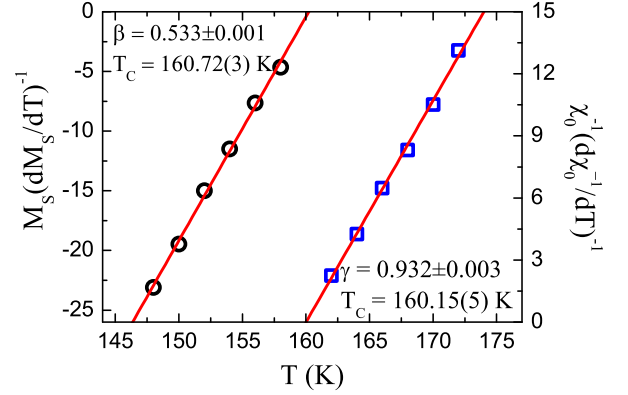}
	\caption{(Color online) Kouvel-Fisher plots associated with spontaneous magnetization $M_S$ (left axis) and inverse initial susceptibility $\chi_0^{-1}$ (right axis) using Eqs. 6 and 7  are shown for nano-crystalline Pr$_2$CoMnO$_6$. Solid lines are due to linear fitting of data.}
	\label{fig:Fig4}
\end{figure}

The $M_S$ and $\chi_0^{-1}$ obtained from intercepts on $M^{1/\beta}$ vs $H/M^{1/\gamma}$ axises by the linear extrapolation from the high field regime in Fig. 2c, have been plotted in Fig. 3. It is clearly seen in Fig. 3 that $M_S$ vs $T$ and $\chi_0^{-1}$ vs $T$ follows Eqs. 1 and 2 reasonably well. The solid lines in Fig. 3 are due to fitting with Eqs. 1 and 2, respectively. The value of exponents $\beta$ = 0.525$\pm$0.005 and $T_c$ = 159.75$\pm$0.05 K obtained from the fitting of $M_S$ with Eq. 1  and $\gamma$ = 0.937$\pm$0.006 and $T_c$ = 160.15$\pm$0.07 K from fitting of $\chi_0^{-1}$ with Eq. 2. The values $\beta$, $\gamma$ and $T_c$ obtained from $M_S$ vs $T$ and $\chi_0^{-1}$ vs $T$ are very much close to the values used in modified Arrott plot in Fig. 2c. The process of determining the values of critical exponents are iterative, where derived critical exponents are used in modified Arrott plot.  The temperature dependent $M_S$ and $\chi_0^{-1}$ in critical region when fitting with equation of state further checks the value of exponents used as trial in modified Arrott plot. 

To further evaluate the values of critical exponents $\beta$, $\gamma$ as well as $T_c$ more precisely we have employed the Kouvel-Fisher (KF) method which is defined as:\cite{kf}

\begin{eqnarray}
  M_S\left(\frac{dM_S}{dT}\right)^{-1} = \frac{(T - T_c)}{\beta} 
\end{eqnarray}

\begin{eqnarray}
  \chi_0^{-1}\left(\frac{d\chi_0^{-1}}{dT}\right)^{-1} = \frac{(T - T_c)}{\gamma}
\end{eqnarray}   

According to above mentioned equations (Eq. 6 and 7) the slopes obtained from plotting of $M_S(dM_S/dT)^{-1}$ vs $T$ yield 1/$\beta$ whereas $\chi_0^{-1}(d\chi_0^{-1}/dT)^{-1}$ vs $T$ gives 1/$\gamma$. Moreover, $T_c$ can be correctly obtained by linear fitting of these plots where intercept on temperature axis gives $T_c$. Kouvel-Fisher plot are shown in Fig. 4 where the $M_S$ and $\chi_0^{-1}$ values obtained from Fig. 4 are plotted using Eq. 5 and 6. The linear fitting of data in Fig. 4 yields $\beta$ = 0.533$\pm$0.001 and $T_c$ = 160.72$\pm$0.03 K and $\gamma$ = 0.932$\pm$0.003 and $T_c$ = 160.15$\pm$0.05 K. The values determined from Kouvel-Fisher method are consistent with the values obtained from modified Arrott plot in Fig. 2c.

So far we have determined two critical exponents $\beta$ and $\gamma$. To determine the third critical exponent  $\delta$ we have used Eq. 3 at critical point $T_c$ = 160 K as discussed above. Fig. 5 shows the plotting of critical isothermal $M(H)$ at $T_c$ = 160 K. Inset Fig. 5 shows the plot of critical isothermal in log-log scale. The $\log M$ vs $\log H$ plot at higher field ($H$$>$ 10 kOe) produce a straight line with slope 1/$\delta$.  Thus the value of third critical exponent from the higher field fitting of straight line in log-log scale yields  $\delta$ = 2.763$\pm$0.005. The obtained value of $\delta$ is close to the theoretical value for mean-field interaction model. Further, according to statical theory these three exponent $\beta$, $\gamma$ and $\delta$ should agree with the Widom scaling equation  which establishes a relationship among these three exponents as following:\cite{widom}

\begin{eqnarray}
  \delta = 1 + \frac{\gamma}{\beta} 
\end{eqnarray}

\begin{figure}
	\centering
		\includegraphics[width=8cm]{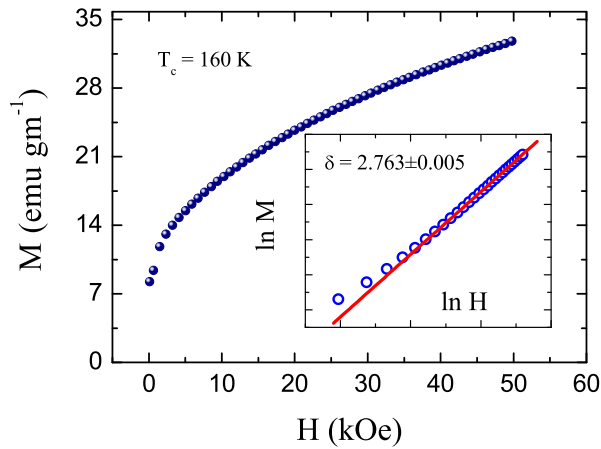}
	\caption{(Color online) The critical isotherm $M(H)$ at $T_c$ = 160 K is plotted for nano-crystalline Pr$_2$CoMnO$_6$. Inset shows log-log plotting of critical isothermal and the solid line is due to linear fitting of data.}
	\label{fig:Fig5}
\end{figure}

Using $\beta$ and $\gamma$ obtained from  Kouvel-Fisher plot we have calculated the value of the exponent $\delta$ = 2.748$\pm$0.005 which is very close to the value obtained from critical isothermal in Fig. 5, further confirms the authenticity of our critical analysis on nano-crystalline Pr$_2$CoMnO$_6$. This proves that the estimation of critical exponents from all methods are consistent. 
The critical exponents ($\beta$, $\gamma$ and $\delta$) and the critical temperature ($T_c$) are tabulated in Table I. Critical exponents determined from different methods exhibit good agreement. It is quite obvious that the exponents are close to mean-field interaction model, yet they do not perfectly mimic the values predicted for any universality class.

\setlength{\tabcolsep}{12pt}
{
\footnotetext[1]{Calculated following Eq. 8}
\begin{table*}[th]
\begin{tabular}{c c c c c c}
\hline
Composition &Ref. &Method &$\beta$ &$\gamma$ &$\delta$\\
\hline
Pr$_2$CoMnO$_6$ &This work &Modified Arrott Plot &0.531 &0.935 &2.748$\footnotemark[1]$\\
 &This work &Kouvel-Fisher Method &0.533$\pm$0.001 &0.932$\pm$0.003 &2.748$\pm$0.003$\footnotemark[1]$\\
 &This work &Critical Isotherm & & &2.763(2)\\
Mean-field Theory & & &0.5 &1.0 &3.0\\
\hline
\end{tabular}
\caption{The values of critical exponents $\beta$, $\gamma$ and $\delta$ determined in this work following various methods for nano-crystalline Pr$_2$CoMnO$_6$ along with the theoretical values of exponents for mean-field model has been tabulated.}
\label{tab:table 1}
\end{table*}
}
 Finally, we have checked the reliability of the values of critical exponents determined for nano-crystalline Pr$_2$CoMnO$_6$ using above discussed self-consistent methods. We have plotted the renormalization magnetization as a function of renormalization field i.e. $M(H,\epsilon)$$\epsilon^{-\beta}$  vs $H$$\epsilon^{-(\beta + \gamma)}$ where the values of exponents used are obtained from Kouvel-Fisher method. The main panel of Fig. 6 shows plotting of Eq. 4, it is evident in the figure that all the data curves collapse into two universal branches one for $T$ $>$ $T_c$ and the other for $T$ $<$ $T_c$. Inset Fig. 6 shows the same plot in logarithmic scale for improved clarity. Thus the data curves satisfied the criteria of scaling hypotheses Eq. 4. Conclusively, it is shown that the critical exponents $\beta$, $\gamma$ and $\delta$ and critical temperature $T_c$ estimated by adopting different methods are quite authentic within the experimental accuracy. This detailed and rigorous investigation confirms that the determined critical exponents and $T_c$ are correct. 

\begin{figure}
	\centering
		\includegraphics[width=8cm]{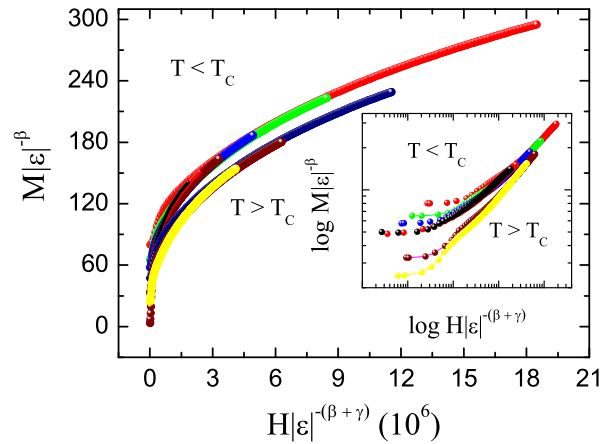}
		\caption{(Color online) Magnetic isotherms collected at both below and above $T_c$ are plotted as renormalized magnetization ($m$) vs renormalized field ($h$) following Eq. 4 for Pr$_2$CoMnO$_6$. The isotherms are collapsed into two branches one below and other above $T_C$.}
	\label{fig:Fig6}
\end{figure}

\subsection{Nature of spin interaction}
There are few established universality classes such as mean field model, 3D Heisenberg model, the 3D Ising model and Tricritical mean field, where FM materials can be classified on the basis of their critical exponents. For nano-crystaline Pr$_2$CoMnO$_6$ the obtained  critical exponents are quite close to the mean field model. To further comprehend the nature of magnetic interaction in nano-crystalline Pr$_2$CoMnO$_6$ we have followed a renormalization group analysis which suggests that spin interaction decays with spatial distance $r$ as; $J(r)$= $r^{-(d+\sigma)}$ where $d$ is the dimensionality of system and $\sigma$ is the range of interaction.\cite{fisher} This model demonstrates that the critical exponent of a particular material reflect the range of magnetic interaction. Further, using the renormalization group approach the exponent for magnetic susceptibility $\gamma$ can be calculated for a particular value of $\left\{d:n\right\}$ as following:\cite{bhatti, fischer}

\begin{eqnarray}
\begin{aligned}
\gamma &= 1 + \frac{4}{d}\frac{(n+2)}{(n+8)} \Delta \sigma + \frac{8(n+2)(n-4)}{d^{2}(n+8)^{2}}\\
&\times \left[1 + \frac{2G(\frac{2}{d})(7n+20)}{(n-4)(n+8)}\right] \Delta\sigma^2
\end{aligned}
\end{eqnarray}

where $\Delta\sigma = (\sigma - \frac{d}{2})$ and $G(\frac{d}{2})= 3 - \frac{1}{4} \left(\frac{d}{2}\right)^2$.  For three dimensional (3D) material ($d$ = 3), thus the exchange interaction can be written as $J(r)$= $r^{-(3+\sigma)}$, for $\sigma$ $\geq$ 2 the Heisenberg model is found valid for 3D isotropic material where $J(r)$ decreases faster then $r^{-5}$. On the other hand if $\sigma$ $\leq$ 1.5 for this condition the mean field model is consider valid where we expect $J(r)$ to decrease more slowly than $r^{-4.5}$. For the understanding of magnetic interaction in understudy compound Pr$_2$CoMnO$_6$, we have substituted the value of $\gamma$ = 0.932 obtained from Kouvel-Fisher method into Eq. 9.  We obtained the value of $\sigma$ = 1.3712 which confirm the long range magnetic interaction for $\sigma$ $\leq$ 1.5 and the $J(r)$ decreases as $J(r)$$\approx$$r^{-4.3712 }$.
This further confirms that the magnetic interaction is long range and mean field model is valid where the exchange coupling decay more slowly than $r^{-4.5}$.

\subsection{Magnetocaloric effect}
Magnetocaloric effect (MCE) characterized by the change in magnetic entropy ($\Delta S_M$) of a material in applied magnetic field $H$ is a vital tool to understand the magnetic phase transition. Thermodynamic Maxwell relation represents how the change in entropy with field is related with change of magnetization with temperature:\cite{phana, tishin}

\begin{eqnarray}
	\left(\frac{\delta S}{\delta H}\right)_T = - \left(\frac{\delta M}{\delta T}\right)_H
\end{eqnarray}

Using classical thermodynamics and  Maxwell relation magnetic entropy change ($\Delta S_M(T,H)$) can be expressed by following relation:

\begin{eqnarray}
	\Delta S_M(T,H) = \int_0^H \left(\frac{\delta M(T, H)}{\delta T}\right)_H dH
\end{eqnarray}

When the magnetization is recorded at small discrete field intervals and at different temperature we can calculate the $\Delta S_M(T,H)$ using following relation:

\begin{eqnarray}
\begin{split}
  \Delta S_M(T,H)\\ 
  = \sum_i \left(\frac{M_{i+1}(T_{i+1}, H) - M_i(T_{i}, H)}{T_{i+1} -T_i}\right) \Delta H
\end{split}
\end{eqnarray}

\begin{figure}[t]
	\centering
		\includegraphics[width=8cm]{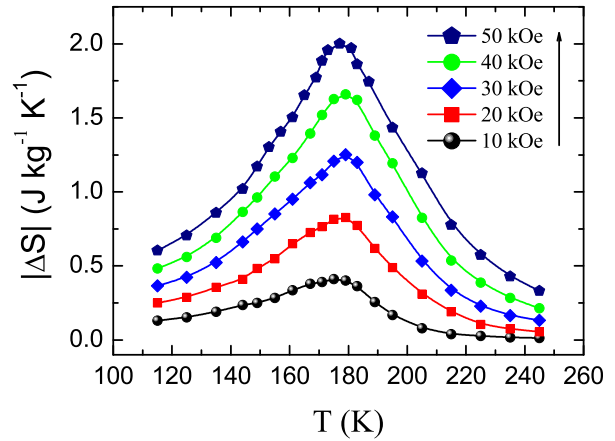}
	\caption{(Color online) The change in magnetic entropy as calculated using Eq. 12 is shown with temperature for nano-crystalline Pr$_2$CoMnO$_6$. Different plots correspond to different highest applied magnetic fields. Lines are guide to eyes.}
	\label{fig:Fig7}
\end{figure}

The change in magnetic entropy ($\Delta S_M$) has been calculated using Eq. 12 for Pr$_2$CoMnO$_6$ from isotherms magnetization $M(H)$ data collected at different temperatures. The $\Delta S_M$ has been calculated for different magnetic fields up to 50 kOe. Fig. 7 shows the $\Delta S_M(T)$ as a function of temperature and magnetic field, the change in entropy shows peak value around PM to FM phase transition.  The calculated $\Delta S_M$ for nano-crystalline Pr$_2$CoMnO$_6$ increases with increasing applied magnetic field with a maximum value of 2.05 J kg$^{-1}$ K$^{-1}$ for 50 kOe applied field.
 The $\Delta S_M$ is not so high to be consider useful for technological applications. However, we have used magnetic entropy change $\Delta S_M$ to understand the nature of magnetic state. According to the mean-field model, the vicinity of a second-order phase transition $\Delta S_M$ shows a power law dependence with scaled field as, $\Delta S_M$ $\propto$ $(H/T_c)^{2/3}$.\cite{oester, hang, bhatti, dong}  Fig. 8a illustrate that  $\Delta S_M$ varies linearly with $(H/T_c)^{2/3}$  which is further consistent with the critical exponent analysis and confirms that the magnetic interaction in nano-crystalline Pr$_2$CoMnO$_6$ follows mean-field interaction model.

It has been shown that for a material which has second order phase transition magnetocaloric data would  follow a scaling model recently proposed by Franco and Conde.\cite{franco} According to this model all the entropy curves $\Delta S_M (T)$ evaluated at different applied magnetic fields after been scaled properly will collapse into a single curve. In this model the normalized entropy change data $\Delta S_M$/$\Delta S_M^{max}$ is plotted as a function of rescaled temperature $\theta$ which is defined as:\cite{franco,thanh, fan}

\begin{eqnarray}
	\theta = \frac{T - T_{pk}}{T_r - T_{pk}}
\end{eqnarray}

where $T_{pk}$ is the peak temperature where $\Delta S_M (T)$ shows maximum value and $T_r$ is the reference temperature  at which $\Delta$S/$\Delta$S$_{M(max)}$ ratio equals 0.8. Fig. 8b shows plotting of $\Delta S_M$/$\Delta S_M^{max}$ vs $\theta$ for different applied magnetic fields. As evident in Fig. 8b, all the entropy change curves, which are significantly different in Fig. 7, collapse into a single line after scaling which confirms the second order nature of phase transition in nano-crystalline Pr$_2$CoMnO$_6$. This further authenticate the critical analysis presented above.

\begin{figure}
	\centering
		\includegraphics[width=8cm]{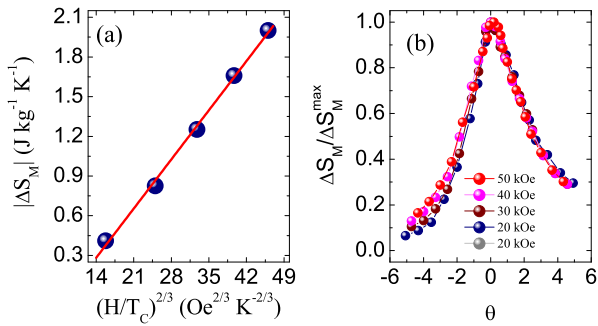}
	\caption{(Color online) (a) Linear dependence of change of magnetic entropy with scaled magnetic field $(H/T_c)^{2/3}$ for Pr$_2$CoMnO$_6$. (b) Shows scaled changes in magnetic entropy data as a function of scaled temperature for nano-crystalline Pr$_2$CoMnO$_6$.}
	\label{fig:Fig8}
\end{figure}

\section{Conclusion} 
In conclusion, we have performed the static scaling in asymptotic region of paramagnetic to ferromagnetic phase transition in nano-crystalline Pr$_2$CoMnO$_6$ is to estimated critical exponents ($\beta$, $\gamma$ and $\delta$) and calculated the change in magnetic entropy. The exponent values are quite close to mean-field model but do not exactly match the values predicted for universality classes. The analysis of exponent using renormalization group approach suggests that spin interaction is of 2-dimensional Heisenberg type having an extended character. We have calculated the change in magnetic entropy using magnetization data. We observe maximum entropy change at 170 K with $\Delta$S = 2.05 J kg$^{-1}$ K$^{-1}$.
 Further, the functional change in magnetic entropy with the applied field obeys the scaling behavior and it is suggestive of mean-field behavior. We conclude that the magnetic ordering in nano-crystalline Pr$_2$CoMnO$_6$ obeys mean field model with long range magnetic interaction.

\section{Acknowledgment}
We acknowledge AIRF (JNU) for measurement facilities. We thank Saroj Jha for the assistance in conducting experiments. Author Ilyas Noor Bhatti acknowledge University Grants Commission, India for financial support.


\begin{thebibliography}{}
\bibitem{moon} J. Y. Moon, M. K. Kim, Y. J. Choi and N. Lee, Scientific Reportsvolume 7, Article number: 16099 (2017).
\bibitem{mahato} R. N. Mahato, K. Sethupathi, V. Sankaranarayanan, R. Nirmala, A. K. Nigam and S. K. Malik Journal of Applied Physics \textbf{109}, 07E319 (2011).
\bibitem{syv} S. Yanez-Vilar, E. D. Mun, V. S. Zapf, B. G. Ueland, J. S. Gardner, J. D. Thompson, J. Singleton,
M. Sanchez-Andujar, J. Mira, N. Biskup, M. A. Senarıs-Rodrıguez and C. D. Batista, Phys. Rev. B \textbf{84}, 134427 (2011).
\bibitem{kumar} S. Kumar, G. Giovannetti, J. van den Brink and S. Picozzi, Phys. Rev. B \textbf{82}, 134429 (2010).
\bibitem{singh} M. P. Singh, K. D. Truong, and P. Fournier, Appl. Phys. Lett. \textbf{91}, 042504 (2007).
\bibitem{gua} H. Guo, J. Burgess, S. Street, A. Gupta, T. G. Calvarese and M. A. Subramanian Appl. Phys. Lett. \textbf{89}, 022509 (2006).
\bibitem{uma} M. Azuma, K. Takata, T. Saito, S. Ishiwata, Y. Shimakawa and M. Takano, J. Am. Chem. Soc. \textbf{127}, 8889 (2005).
\bibitem{yi} W. Yi, A.J. Princep, Y. Guo, R. D. Johnson, D. Khalyavin, P. Manuel, A. Senyshyn, I. A. Presniakov, A. V. Sobolev, Y. Matsushita, M. Tanaka, A. A. Belik and A. T. Boothroyd, Inorg. Chem. \textbf{54}, 8012 (2015).
\bibitem{sco1} J. Blasco, J. García, G. Subías, J. Stankiewicz, J. A. Rodríguez-Velamazán, C. Ritter, J. L. Garcí -Muñoz, and F. Fauth, Phys. Rev. B \textbf{93}, 214401 (2016).
\bibitem{sco2} J. Blasco, G. Subías, J. Garcia, J. Stankiewicz, J. A. Rodriguez-Velamazán, C. Ritter, and J. L. Garcia-Munoz, Solid State Phenomena \textbf{257}, 95 (2017).
\bibitem{zli} Z. Li, Y. Cho, XXiang Li, Xinyu Li, A. Aimi, Y. Inaguma, J. A. Alonso, M. T. Fernandez-Diaz, J. Yan, M. C. Downer, G. Henkelman, J. B. Goodenough and J. Zhou, J. Am. Chem. Soc. \textbf{140}, 2214 (2018).
\bibitem{jsu} J. Su, Z. Z. Yang, X. M. Lu, J. T. Zhang, L. Gu, C. J. Lu, Q. C. Li, J.-M. Liu and J. S. Zhu, ACS Appl. Mater. Interfaces \textbf{7}, 13260 (2015).
\bibitem{rov} I.N. Flerov, M. V. Gorev, K.S. Aleksandrov, A. Tressaud, J. Grannec and M. Couzi, Mater. Sci. Eng. \textbf{24}, 81 (1998).
\bibitem{dass} R. I. Dass and J. B. Goodenough, Phys. Rev. B \textbf{67}, 014401 (2003).
\bibitem{son} M. Anderson, K. B. Greenwood, G. A. Taylor, and K. R. Poeppelmeier, Prog. Solid State Chem. \textbf{22}, 197 (993).
\bibitem{dya} S. Baidya and T. Saha-Dasgupta, Phys. Rev. B \textbf{84}, 035131 (2011).
\bibitem{choudhury} D. Choudhury, P. Mandal, R. Mathieu, A. Hazarika, S. Rajan, A. Sundaresan, U. V. Waghmare, R. Knut, O. Karis, P. Nordblad, and D. D. Sarma, Phys. Rev. Lett. 108:127201, (2012).
\bibitem{kawa}  Y. Shimakawa, M. Azuma, and N. Ichikawa, Materials \textbf{4}, 153 (2011).
\bibitem{lil} H. Nhalil, H. S. Nair, C. M. N. Kumar, A. M. Strydom and S. Elizabeth, Phys. Rev. B \textbf{92}, 214426 (2015).
\bibitem{bos} J-W G. Bos and J. P. Attfield, Phys. Rev. B \textbf{70}, 174434 (2004).
\bibitem{ala} S. Vasala and M. Karppinen, Progress in Solid State Chemistry \textbf{43},  01 (2015).
\bibitem{maya} M. Retuerto, A. Munoz, M. J. Martínez-Lope, J. A. Alonso, F. J. Mompean, M. T. Fernández-Díaz, and J. Sanchez-Benítez Inorg. Chem. \textbf{54}(22), 10890 (2015).
\bibitem{lshi} L. Shi, W. Liu, J. Zhao, Y. Li, S. Zhou, Y. Guo and Y. Wang, Mater. Res. Express \textbf{2} 076104 (2015).
\bibitem{shi} R. Takahashi, I. Ohkubo, K. Yamauchi, M. Kitamura, Y. Sakurai, M. Oshima, T. Oguchi, Y. Cho and M. Lippmaa, Phys. Rev. B. \textbf{91}, 134107 (2015).
\bibitem{liu} W. Liu, L. Shi, S. Zhou, J. Zhao, Y. Li and Y. Guo, J. Appl. Phys. \textbf{116}, 193901 (2014).
\bibitem{wang} X. L. Wang, J. Horvat, H. K. Liu, A. H. Liv and S. X. Dou, Solid State Communications \textbf{118}, 27 (2001.)
\bibitem{ilyas} I. N. Bhatti, R. N. Mahato, I. N. Bhatti and M. A. H. Ahsan, arXiv:1810.03895 [cond-mat.str-el].
\bibitem{khemchand} K. Rawat, Meenakshi and R. N. Mahato, Mater. Res. Express 5,066110 (2018).
\bibitem{zhang} X. Zhang, S. Matsuishi, and H. Hosono, J. Phys. D: Appl. Phys. \textbf{49}, 335002 (2016).
\bibitem{michael} Michael E. McHenry and David E. Laughlin, Theory of Magnetic Phase Transition in \textit{Characterization of Materials}, 1st ed., \textbf{1}, pp. 528–530, Carnegie Mellon University, Pittsburgh, Pennsylvania (2002).
\bibitem{stanley} H. E. Stanley, \textit{Introduction to phase transitions and critical phenomenon}, Oxford University Press, New York, (1971).
 \bibitem{bee} S.K. Banerjee, Phys. Lett. \textbf{12}, 16 (1964).
\bibitem{arrott} A. Arrott, Phys. Rev. \textbf{108}, 1394 (1957).
\bibitem{kf} J. S. Kouvel and M. E. Fisher, Phys. Rev. \textbf{136}, A1626 (1964).
\bibitem{arrott1} A. Arrott and J. E. Noakes, Phys. Rev. Lett. \textbf{19}, 786 (1967).
\bibitem{widom} B. Widom, J. Chem. Phys. \textbf{43} (1965), 3898;J. Chem. Phys. \textbf{41}, 1633 (1964).
\bibitem{fisher} M. E. Fisher, Shang-keng Ma, and B. G. Nickel, Phys. Rev. Lett. \textbf{29}, 917 (1972).
\bibitem{bhatti}	I. N. Bhatti and A. K.  Pramanik, J. Magn. Magn. Mater. \textbf{422}, 141 (2017).
\bibitem{fischer} S. F. Fischer, S. N. Kaul, and H. Kronm$\ddot{u}$ller, Phys. Rev. B. \textbf{65}, 064443  (2002).
\bibitem{phana} M. H. Phana and S. C. Yu, J. Magn. Magn. Mater. \textbf{308}, 325  (2007).
\bibitem{tishin} A. M. Tishin and I. Spichkin, \textit{The Magnetocaloric Effect and Its Applications} Institute of Physics, Bristol, (2003). 
\bibitem{oester} H. Oesterreicher and F. T. Parker, J. Appl. Phys. \textbf{55}, 4334 (1984).
\bibitem{hang} X. Zhang, S. Matsuishi and H. Hosono, J. Phys. D: Appl. Phys. \textbf{49} 335002  (2016).
\bibitem{dong} Q. Dong, H. Zhang, J. Shen, J. Sun and B. Shen, J. of Magn. Magn. Mater. \textbf{319}, 56 (2007).
\bibitem{franco} V. Franco and A. Conde, Int. J. Refrig. \textbf{33},  465 (2010).
\bibitem{fan} J. Fan, L. Pi, L. Zhang, W. Tong, L. Ling, B. Hong, Y. Shi, W. Zhang, D. Lu, and Y. Zhang, Appl. Phys. Lett. 98, 072508 (2011). 
\bibitem{thanh} T. D. Thanh, D. C. Linh, T. V. Manh, T. A. Ho, T-L Phan, and S. C. Yu, J. Appl. Phys. \textbf{117}, 17C101 (2015).
\end{thebibliography}
\end{document}